\documentclass{PoS}

\usepackage{color}
\usepackage{amsmath}
\usepackage{graphicx}
\usepackage{amssymb}
\usepackage{esint}

\def\R{\text{\rm R}}

\def\e{{\rm e}}

\def\s{\sigma}
\def\S{\Sigma}
\def\nn{\nonumber}
\def\p{\partial}
\def\ls{\left[}
\def\rs{\right]}
\def\lc{\left\{}
\def\rc{\right\}}
\newcommand{\be}{\begin{eqnarray}}
\newcommand{\ee}{\end{eqnarray}}

\title{Quintessence from higher curvature supergravity }

\ShortTitle{Higher curvature supergravity}

\author{
\speaker{
Fotis Farakos
}
\\
        KU Leuven, Institute for Theoretical Physics, 
			Celestijnenlaan 200D, B-3001 Leuven, Belgium\\
        E-mail: \email{fotios.farakos@kuleuven.be} }

\abstract{
In this contribution we revisit higher curvature N=1 supergravity and discuss the quintessence phase that can appear due to the $R^4$ terms. 
In particular we focus on the bosonic supersymmetric completion within the old-minimal and the new-minimal formulations. 
}

\FullConference{Corfu Summer Institute 2019 "School and Workshops on Elementary Particle Physics and Gravity" (CORFU2019)\\
		31 August - 25 September 2019\\
		Corfu, Greece}

\begin{document}

\def\ls{\left[}
\def\rs{\right]}
\def\lc{\left\{}
\def\rc{\right\}}

\def\p{\partial}

\def\S{\Sigma}

\def\s{\sigma}

\def\O{\Omega}

\def\a{\alpha}
\def\b{\beta}
\def\g{\gamma}

\def\ad{{\dot \alpha}}
\def\bd{{\dot \beta}}
\def\gd{{\dot \gamma}}

\def\nn{\nonumber}

\section{Introduction and discussion}

The exact nature of the dark energy in our universe is currently lacking 
a concluding explanation both in theoretical and in observational physics. 
From the observational physics perspective, 
a cosmological constant of the order $\Lambda \sim 10^{-120}$ (in Planck units) 
is a perfectly valid explanation for the accelerated expansion of our universe, 
but at the same time one cannot exclude an underlying quintessence phase, 
i.e. a slightly time-varying $\Lambda$. 
From a theoretical perspective, 
at this stage there is also no consensus on the capacity of string theory 
to provide de Sitter vacua, 
see e.g. \cite{Danielsson:2018ztv,Obied:2018sgi,Kallosh:2018nrk}\footnote{For more recent developments 
see for example \cite{Cordova:2018dbb,Blaback:2019zig,Cribiori:2019clo}.}. 
Therefore the quintessence type of dark energy \cite{Wetterich:1987fm,Ratra:1987rm,Caldwell:1997ii,Tsujikawa:2013fta}, 
that can be provided by runaway potentials, 
deserves a detailed study both in supergravity 
and in string theory 
\cite{Brax:1999gp,Copeland:2000vh,Hellerman:2001yi,Chiang:2018jdg,Cicoli:2018kdo,
Olguin-Tejo:2018pfq,Emelin:2018igk,Farakos:2019ajx,Ferrara:2019tmu,Hebecker:2019csg,DallAgata:2019yrr,Montero:2020rpl}. 
In this contribution we will study quintessence from the perspective of a pure gravitational theory, 
we will show how such dynamics can naturally arise, 
and we will in particular embed these models in four-dimensional N=1 supergravity.

Higher curvature gravitation of the form ${\cal F}(R)$, 
where $R$ is the Ricci scalar, 
has the property to provide an additional scalar degree of freedom, 
the so-called scalaron, 
which will generically have a canonical kinetic term and will also have a scalar potential \cite{Whitt:1984pd}. 
This property of higher curvature gravitation is utilized in the 
Starobinsky model of inflation \cite{Starobinsky:1980te} which is simply given by the Lagrangian 
\be
\label{RR2}
- \frac12 \sqrt{- g} M_P^2 R + \frac{M_P^2}{12 m^2} \sqrt{- g} R^2 \, , 
\ee
and is described equivalently by a scalar-tensor theory with scalar potential given by 
\be
V_{R+R^2} = \frac34 m^2 M_P^2 \left( 1 - \e^{- \sqrt{\frac23} \phi / M_P} \right)^2 \, .  
\ee
Here $\phi$ is the real scalar that is identified as the scalaron. 
For large values of $\phi$ the model describes an inflationary phase. 
We now make the observation that adding the term 
\be
+ \xi \sqrt{- g} R^4 
\ee
in the Lagrangian density \eqref{RR2}, 
leads to a scalar potential with runaway behavior for large values of $\phi$. 
Indeed, 
for large values of $\phi$ the scalaron potential behaves as 
\be
\label{RAW}
V_{+R^4}|_{\phi \gg M_P} \sim  \xi^{-1/3} \, \e^{- \frac{2}{3} \sqrt{\frac23} \phi / M_P} \, . 
\ee  
As a result we see that $R^4$ gravity can provide a viable quintessence model that gives 
\be
0 < M_P |V' / V|_{\phi \gg M_P} < 0.6 \, , 
\ee 
and it is in agreement both with the cosmological observational data and 
with the late-time cosmological implications of the de Sitter conjectures \cite{Agrawal:2018own}. 
The asymptotic behavior in \eqref{RAW} is derived by assuming that the term 
$\e^{\sqrt{\frac23} \phi / M_P}$ always dominates. 
To take the limit $\xi \to 0$ one has to consider the complete form of the scalar potential \eqref{RAW}, 
which can be found for example in \cite{Farakos:2015ksa} and is depicted here in figure 1.

\begin{figure}[htp] \centering{
\includegraphics[scale=0.72]{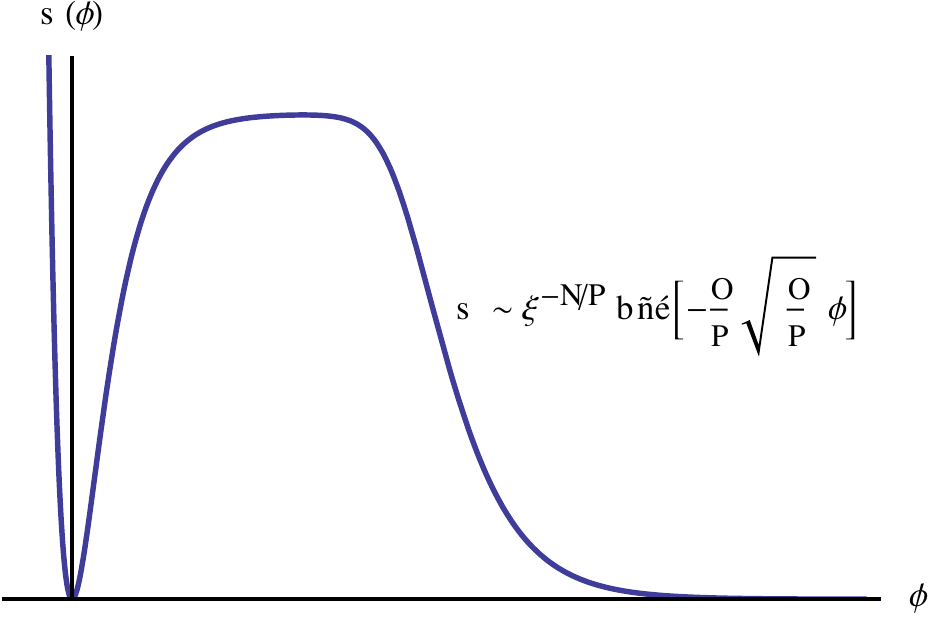}}
\caption{The potential $V(\phi)$ and its asymptotic behavior for large values of the scalaron $\phi$.} 
\end{figure}

The interest in embedding these gravitational type of runaway models in minimal supergravity is their universality. 
Indeed, we can draw generic conclusions for the runaway phase, 
if such phase originates from the pure supergravity multiplet sector 
and if the impact of the matter couplings (and other supersymmetry breaking sectors) 
on the background dynamics can be ignored. 
Under these assumptions we study the simplest two minimal 4D N=1 supergravity formulations: 
the old-minimal (see e.g. \cite{Wess:1992cp,FVP}) and the new-minimal (see e.g. \cite{Ferrara:1988qxa,Ferrara:2018dyt}). 
As we will see these two formulations lead to distinct phenomenological implications related to the fact that the 
runaway scalar resides either in a massive vector multiplet \cite{VanProeyen:1979ks,Cecotti:1987qe} (for new-minimal) 
or a massive double-chiral multiplet \cite{Ferrara:1978rk,Theisen:1985jr,Cecotti:1987sa} (old-minimal supergravity).

In the next two sections we study the structure of the old-minimal higher curvature supergravity 
and the properties of the quintessence phase, 
whereas in the fourth section we study the higher curvature terms for the new-minimal supergravity. 
We generically set $M_P=1$ unless when it explicitly appears. 
We refer the reader to \cite{Wess:1992cp} for conventions.

\section{Old-minimal $R+R^2$ supergravity}

The old-minimal formulation of four-dimensional N=1 supergravity contains the vierbein $e_m^{\ a}$, 
the gravitino $\psi_m^{\ \alpha}$, 
the complex scalar auxiliary field $M$ and the real vector auxiliary field $b_a$. 
The free supergravity Lagrangian is constructed by the 
Lagrangian superspace density 
\be
\label{SR}
{\cal L}_{R} = - 3 \int d^2 \Theta \, 2 {\cal E} \, {\cal R} + c.c. 
\ee 
The bosonic contribution of the chiral density $2 {\cal E}$ is given by 
$2{\cal{E}}=e - \Theta \Theta e \overline M$. 
The Ricci superfield ${\cal R}$ is chiral, i.e. $\overline{ {\cal D}}_{\dot{\alpha}} {\cal R} =0$, 
and its 
lowest component is given by the auxiliary field $M$, namely ${\cal R} | = - M /6 $, 
where we use the abbreviation $``|"$ standing for $|_{\theta=\overline \theta =0}$. 
For our discussion the other relevant component of ${\cal R}$ is the highest component that has bosonic contributions 
\be
{\cal D}^2 {\cal R} | =  -\frac{1}{3} R + \frac{4}{9} M \overline M + \frac{2}{9} b^a b_a 
-\frac{2 i}{3} e_{a}^{\ m} D_m b^a \, . 
\ee
Here $D_m$ is the covariant derivative for Lorentz indices and the connection it contains is the $\omega_{klm}$. 
Using these ingredients the bosonic sector of \eqref{SR} takes the form 
\be
\label{OML} 
e^{-1} {\cal L}_{R} =  - \frac{1}{2} R -\frac{1}{3} M \overline M + \frac{1}{3} b^a b_a \, . 
\ee
From \eqref{OML} we see that the auxiliary fields are integrated out to give 
\be
M =0 \ , \quad b_a = 0 \, . 
\ee 
Inserting these values back into \eqref{OML} gives the standard N=1 supergravity Lagrangian.

From the structure of the chiral superfield ${\cal R}$ we see that a superspace term of the form 
$\int d^4 \theta E \, {\cal R} \overline{\cal R}$ will generate a component term of the form $e\,R^2$, 
which contains the scalaron \cite{Whitt:1984pd}. 
However, 
on top of the scalaron, 
the aforementioned superspace term will also generate kinetic terms for the scalar $M$ and for the 
scalar $\nabla_m b^m$ \cite{Ferrara:1978rk}. 
As a result the $R^2$ supergravity will contain 4 real scalar degrees of freedom which are known to fall into 
two chiral multiplets $T$ and $S$ \cite{Cecotti:1987sa}. 
The exact classical equivalence between the two descriptions is known. 
In particular, 
the generic form of an $R^2$ supergravity action can be described by the superspace term \cite{Theisen:1985jr,Hindawi:1995qa} 
\be
\label{R2}
{\cal L}_{R+R^2} =  \frac{3}{8} \int d^2 \Theta \,  2 {\cal E} \,  (\overline {\cal D}^2 - 8 {\cal R} )  f({\cal R},\overline {\cal R}) + c.c. 
\ee 
Using their classical equivalence, 
the Lagrangian \eqref{R2} can be written as standard supergravity 
\be
\label{R2eq}
{\cal L}_{R+R^2} = \Big{[} \frac{3}{8} \int d^2 \Theta \,  2 {\cal E} \,  (\overline {\cal D}^2 - 8 {\cal R} )  
\e^{- \frac{1}{3} K} 
+ \int d^2 \Theta \,  2 {\cal E} \, W \Big{]} +c.c. \, , 
\ee
with K\"ahler potential 
\be
\label{KR2}
K = - 3 \, \text{ln} \lc  T  + \overline T + f(S,\overline S) \rc \, , 
\ee
and superpotential 
\be
\label{WR2}
W = 6 \, T S  \, . 
\ee
The duality procedure that relates \eqref{R2} to \eqref{R2eq} has been presented in detail in \cite{Cecotti:1987sa} 
in the superconformal setup and in \cite{Dalianis:2014aya} in the framework of Poincar\'e superspace, 
therefore we will not review it here.

The duality works for any form of the function $f({\cal R},\overline {\cal R})$ in \eqref{R2}. 
We will however restrict our discussion to a specific form for the function $f(S, \overline S)$ such that the R-symmetry is preserved. 
In particular we choose the form 
\be
\label{fS} 
f(S, \overline S) = 1 - 2  \frac{ S \overline S}{m^2}  
+ \frac19 \zeta \, \frac{S^2 \overline S^2}{m^4} \, , 
\ee
which appears in the standard formulation of the Starobinsky model in four-dimensional supergravity \cite{Kallosh:2013lkr,Farakos:2013cqa}. 
With this form of the function 
the complex scalar residing in $S$  will always be strongly stabilized 
to 
$s = 0$ ($s=S|$) and can be made arbitrarily heavy 
for large values of $t$ ($t =$Re$\,T|$) depending on the parameter $\zeta$ \cite{Dalianis:2014aya,Kallosh:2013lkr}. 
The imaginary part of the lowest component of $T$, 
also becomes massive and gets strongly stabilized at $b=0$ ($b=$Im$\,T|$) because of the curvature of the scalar manifold. 
For large values of $t$ notice that 
supersymmetry is always spontaneously broken by $\langle F^S \rangle \ne 0$ 
and the goldstino is aligned with the fermion of the chiral superfield $S$. 
As a result, 
one can use instead of $S$ the nilpotent chiral superfield $X$ ($X^2=0$) \cite{Antoniadis:2014oya}, 
as long as the $t$ does take large values \cite{DallAgata:2014qsj}. 
Finally, 
in the standard embedding of the Starobinsky model in supergravity the scalar potential 
develops a plateau for large values of $t$ and inflation takes place with the inflationary scale 
given roughly by $H^2 \sim m^2$. 
Further cosmological aspects of higher curvature supergravity systems have been analyzed 
for example in \cite{Dalianis:2014aya,Terada:2014uia,Dalianis:2015fpa,Hasegawa:2015era,Dalianis:2018afb}.

\section{Including the $R^4$ terms}

In the standard embedding of the Starobinsky model in N=1 supergravity the inflationary trajectory ends 
in a Minkowski supersymmetric vacuum at $T=0$, $S=0$. 
We will now show that once $R^4$ terms are taken into account the theory develops naturally a 
quintessence behavior for large values of $t$, 
and we will study the stability of the supergravity theory at that 
phase.\footnote{Note that not all supersymmetrizations of $R^4$ terms are the same in old-minimal supergravity. 
In particular the $R^4$ supersymmetrization presented in \cite{Cecotti:1987sa} 
(and later in \cite{Moura:2002ft}) will contain additional negative-norm states.}

We first introduce the appropriate $R^4$ term, 
which in old-minimal supergravity has the form 
\be
\label{SGR4}
{\cal L}_{R^4} = - 3^4 \xi \int d^2 \Theta \, 2 {\cal E} \, (\overline {\cal D}^2 - 8 {\cal R} ) \Big{|} {\cal D} {\cal R}  \Big{|}^4  + c.c. 
\ee
where 
\be
\label{mimimimi}
\Big{|} {\cal D} {\cal R}  \Big{|}^4 = {\cal D}^\alpha {\cal R} \, {\cal D}_\alpha {\cal R} \, 
\overline{\cal D}_{\dot \alpha} \overline{\cal R} \, 
\overline{\cal D}^{\dot \alpha} \overline{\cal R} \, . 
\ee
Once we project to components we find ${\cal L}_{R^4} =e \xi R^4 + \dots$ 
Following \cite{Farakos:2013cqa} we will study this theory directly in the dual form that is given by 
replacing ${\cal R}$ with $S$ in \eqref{SGR4} (and \eqref{mimimimi} equivalently). 
The steps to perform this duality are exactly the same as the ones described in detail in \cite{Dalianis:2014aya,Farakos:2013cqa}. 
Therefore the complete Lagrangian we will study has the from 
\be
\label{TOT1}
\begin{aligned}
{\cal L}_\text{TOT} = & \Big{[} \frac{3}{8} \int d^2 \Theta \,  2 {\cal E} \,  (\overline {\cal D}^2 - 8 {\cal R} )  
\e^{- \frac{1}{3} K} 
+ \int d^2 \Theta \,  2 {\cal E} \, W \Big{]} +c.c. \, 
\\ & - 3^4 \xi \int d^2 \Theta \, 2 {\cal E} \, (\overline {\cal D}^2 - 8 {\cal R} ) \Big{|} {\cal D} S  \Big{|}^4  + c.c. 
\end{aligned}
\ee
with $K$ and $W$ given by \eqref{KR2} and \eqref{WR2}, 
and $| {\cal D} S |^4 = {\cal D}^\alpha S \, {\cal D}_\alpha S \, 
\overline{\cal D}_{\dot \alpha} \overline S \, 
\overline{\cal D}^{\dot \alpha} \overline S$. 
For the complete bosonic sector of this Lagrangian see \cite{Farakos:2013cqa}. 
For further properties of the $| {\cal D} S |^4$ terms 
see \cite{Koehn:2012ar,Farakos:2012qu,Kamada:2014gma,Cicoli:2016chb,Weissenbacher:2019bfb}. 
The scalars are stabilized at\footnote{We refer to the lowest component of a chiral superfield with the same letter as for the chiral superfield itself.}  
\be
\label{BG}
\langle S \rangle = 0 \, , \quad \langle \text{Im} T \rangle = 0 \, , 
\ee
and the kinetic term for the Re$T$ is canonically normalized once we set 
\be
\text{Re} T = \frac12 \e^{\sqrt{\frac23} \phi / M_P} - \frac12  \, . 
\ee 
We will not give a full expression for the kinetic terms or the scalar potential here, 
rather we will bring forward the important results for our discussion.

For the background \eqref{BG}, 
the form of the scalar potential for large values of $\phi$ is 
\be
\label{asympt}
V|_{\phi \gg M_P} \sim \xi^{-1/3} \, \e^{- \frac{2}{3} \sqrt{\frac23} \phi / M_P} \, , 
\ee  
and the system can describe a quintessence phase. 
From the expectation values of the fields on the runaway background we see that the R-symmetry is preserved 
which means that the Lagrangian gravitino mass is vanishing. 
As a result, 
the vacuum energy is also identified with the supersymmetry breaking scale, 
that is 
\be 
\label{fsusy}
\langle m_{3/2} \rangle=0 \, \to \,  f_{SUSY} = \sqrt{V|_{\phi \gg M_P}} \, ,  
\ee
and predicts a very low supersymmetry breaking scale. 
Notice that the prediction of this model for the c-parameter of \cite{Obied:2018sgi} is 
independent of the scales that enter the theory and reads 
\be
c = M_P |V' / V| = \frac{2}{3} \sqrt{\frac23} \sim 0.54 \,. 
\ee
Finally the mass of $\phi$ is directly related to the supersymmetry breaking scale 
\be
m_\phi^2 \sim \e^{-\frac{2}{3} \sqrt{\frac23} \phi} \xi^{-\frac13} \sim f_{SUSY}^2 / M_P^2 \, . 
\ee

Let us now turn to the other scalar masses and examine the stability of the runaway phase. 
The kinetic terms of the scalars $S$ and Im$T$ are not canonical, therefore one has to first canonically normalize them. 
If we study only quadratic fluctuations this is easily done. 
For example if we have a scalar $\chi$ with fluctuations 
\be
-\frac12 {\cal K}(\phi) (\p \chi)^2 - \frac12 {\cal M}^2(\phi) \chi^2 \, , 
\ee
then the effective canonically normalized mass is simply $m^2_{\chi-{\rm can.}} = {\cal M}^2(\phi)/{\cal K}(\phi)$. 

From the component form of the Lagrangian \eqref{TOT1}, 
the mass of the complex scalar $S$, 
once it is canonically normalized, 
is up to a numerical factor given by 
\be 
M_{S-\text{can.}}^2 \sim m^2 \Big{[} \frac{\zeta \, \e^{-\sqrt{\frac23} \phi}}{(2^6 3^{9/2} m^6 \xi)^{2/3}} - 24 \, \e^{- \frac23 \sqrt{\frac23} \phi} \Big{]} \, . 
\ee 
Here three comments are in order. 
Firstly, 
when we evaluate the kinetic term of $S$ in order to canonically normalize it, 
we find that there are two contributions: 
The contribution from the standard kinetic term related to the K\"ahler potential 
and a contribution from the higher derivative terms (not ghost however) 
that originate from the last line in \eqref{TOT1}. 
The latter in fact dominate over the former. 
Secondly, the mass of the scalar $S$ is very small as it is suppressed with exponential factors of $\phi$. 
Thirdly, the mass term has one positive and one negative contribution. 
The positive contribution originates from the stabilizer term proportional to the $\zeta$ parameter in the K\"ahler potential. 
Clearly one needs an unnaturally large value for $\zeta$ to have a stable $S$, 
otherwise the system will suffer from a tachyonic instability due to negative $M_S^2$. 
It is interesting to notice that 
while $\phi$ grows 
so does the negative contribution to the mass. 
Therefore the $M_S^2$ will unavoidably, 
at some late time, 
become negative once $\phi$ goes beyond a critical value and the system will collapse. 
Such value of course might be outside the regime of validity of the supergravity effective field theory, 
or may signal that the effective theory breaks down.

Finally for the imaginary component of the complex scalar $T$, 
namely Im$T = b$,  
we have once it is canonically normalized (up to numerical factors) 
\be
M_{b-\text{can.}}^2 \sim \e^{-\frac{2}{3} \sqrt{\frac23} \phi} \xi^{-\frac13} \sim f_{SUSY}^2 / M_P^2 \, .   
\ee
We conclude that the scalar $b$ will be again very light. 
Notice that the mass and the couplings of $b$ are uniquely determined by the higher curvature supergravity. 
Clearly a signal of this setup will be two very light scalars ($\phi$ and $b$) with almost degenerate masses of the order 
of the observable Hubble scale $\text{H}_\text{obs.}$.

Before closing our discussion for the old-minimal supergravity let us return to the observation 
that the model we presented has a vanishing Lagrangian gravitino mass. 
There are two way to deal with such an issue. 
One way is to add a constant superpotential $W_0$. 
This however destabilizes the system away from $b=0$ and makes it very hard to find a tracktable quintessence phase. 
An alternative direction is to introduce an explicit gravitino mass term of the form presented in \cite{Farakos:2017mwd}, 
where the role of $X$ is taken here by $S$. 
By introducing these terms one effectively also changes the scale of supersymmetry breaking, 
which now should be evaluated as ${\cal F} = \sqrt{V + 3 m_{3/2}^2}$. 
This on one hand makes the phenomenology more attractive, 
but on the other hand it is very unclear how such gravitino mass terms could consistently arise from string theory.

\section{New-minimal supergravity}

Now we turn to the new-minimal four-dimensional supergravity. 
One can easily see that the higher curvature theory will not have moduli because the scalaron supermultiplet 
is in this setup a massive vector multiplet \cite{VanProeyen:1979ks,Cecotti:1987qe,Farakos:2013cqa}. 
We will follow the setup presented in \cite{Farakos:2015ksa} 
and see that quintessence models also arise in a straightforward manner in this case.

The new-minimal supergravity contain as propagating degrees of freedom the graviton and the gravitino, 
and it contains the gauge field for the R-symmetry $A_m$ and the gauge two-form $B_{mn}$ as auxiliary fields. 
Comprehensive reviews of the structure of this theory can be found in \cite{Ferrara:1988qxa,Ferrara:2018dyt}, 
therefore we will not review it here, 
rather we will directly bring forward the important ingredients - we will mostly follow \cite{Farakos:2015ksa}. 
The central multiplet here is the $U(1)_\R$ gauge multiplet $V_\R$. 
It has component field 
\be
\begin{aligned}
- \frac12 [{\cal D}_\alpha , \overline{\cal D}_{\dot \alpha} ]  V_\R | & = A^-_{\alpha \dot \alpha} = A_{\alpha \dot \alpha} - 3 H_{\alpha \dot \alpha} \, , 
\\
\frac18 {\cal D}^\alpha \overline{\cal D}^2 {\cal D}_\alpha V_\R | & = - \frac12 \left( -R + 6 H^aH_a \right) \, , 
\end{aligned}
\ee
where $D^a H_a=0$, 
therefore $H_m$ is the Hodge-dual of the auxiliary field two-form $B_{mn}$. 
We are interested in a Lagrangian that reproduces the $R^4$ term in the bosonic sector, 
but includes the appropriate couplings for its supersymmetric completion. 
We have 
\be
{\cal L}_{\rm TOT} = - 2 \int d^4 \theta E \, V_\R + \left\{ \frac{\alpha}{4} \int d^2 \theta \, {\cal E} \, W^2 + c.c. \right\}
 + 16 \xi \int d^4 \theta E \, W^2 \overline W^2 \, ,  
\ee
where $W_\alpha= W_\alpha(V_\R) = - \frac14 \overline{\cal D}^2 {\cal D}_\alpha V_\R$. 
Here the first term describes the Hilbert--Einstein term, 
the second term describes $\alpha R^2$ and the last term describes $\xi R^4$. 
The full bosonic sector reads 
\be
\begin{aligned}
e^{-1} {\cal L}_{\rm TOT}  = &  - \frac12 R + 2 A^a H_a -3 H^a H_a 
+  \frac{\alpha}{8} \left( R - 6 H^2 \right)^2 - \frac{\alpha}{4} F^2(A^-) 
\\
&+  \xi \left[ 2 F^2(A^-) - (R - 6 H^2)^2 \right]^2 
+  \xi \left( F^{kl}(A^-) F^{mn}(A^-) \epsilon_{klmn} \right)^2 
+ 2 X D^a H_a \, , 
\end{aligned}
\ee
where in the last line we have also introduced a Lagrange multiplier $X$ that will enforce $D \cdot H=0$ once integrated out. 
Now we redefine the gauge vector as 
\be
v_m = A_m - 3 H_m - \partial_m X \, , 
\ee
and we have to introduce two Lagrange multipliers $Z$ and $Y$ to bring the system to a linear form. 
The classically equivalent Lagrangian we have to use is 
\be
\label{NM1}
\begin{aligned}
e^{-1} {\cal L}_{\rm EQ}  = &  - \frac12 R + 2 v^a H_a +3 H^a H_a 
\\
& +  \frac{\alpha}{8} Y^2 
- \frac{\alpha}{4} F^2 
+  \xi \left( F^{kl} F^{mn} \epsilon_{klmn} \right)^2 
\\
& +  \xi \left( 2 F^2 - Y^2 \right)^2 
- Z \left[ R - 6 H^2 + Y \right] \, , 
\end{aligned}
\ee
where now $F_{mn}=\p_m v_n - \p_n v_m$. 
The Lagrangian \eqref{NM1} has still many non-linear terms but we are interested in two basic properties: 
How it describes quintessence and what is the mass of the vector. 
It is straightforward to derive the scalar potential for the quintessence phase but we will do so by focusing on the 
steps that are relevant for finding also the vector mass.

First we notice that by setting to vanish all the terms that contribute to the $v_m$ sector we have 
\be
\label{NM2}
e^{-1} {\cal L}_{Grav.} =  - \left( Z+ \frac12 \right) R +  \frac{\alpha}{8} Y^2 - Z Y +  \xi Y^4  \, . 
\ee
To find the effective scalar-gravity theory we have to integrate out $Y$ which gives a cubic equation 
of the form 
\be
Y^3 + \frac{\alpha}{16 \xi} Y - \frac{Z}{4 \xi} = 0 \ \to \ Y(Z) \, ,  
\ee 
which once solved and inserted back into the action \eqref{NM2} will give a scalar potential  for $Z$, $V(Z)$. 
Notice that for large $Z$ values we have 
\be
Y(Z)\Big{|}_{{\rm large} \, Z} \to \left(  \frac{Z}{4 \xi} \right)^{1/3}\, .  
\ee
However the theory will still be in Jordan frame. 
To go to the Einstein frame we perform a Weyl rescaling with 
\be
g_{mn} \to  g_{mn} \left( Z+ \frac12 \right)^{-1} \, . 
\ee
In the end we find a scalar potential for the canonically normalized scalar $\phi$, 
which is found by setting 
\be
Z = \frac12 \e^{\sqrt{\frac23} \phi} - \frac12 \, . 
\ee
For large $\phi$ values the scalar potential takes exactly the form \eqref{RAW} and gives quintessence. 
Further details for the exact procedure can be found in \cite{Farakos:2015ksa}. 
Further discussions on the cosmological properties of a massive vector multiplet 
can be found in \cite{Ferrara:2013rsa,Ferrara:2013kca,Addazi:2017ulg,Aldabergenov:2020pry}.

Let us now turn to the massive vector. 
We are interested only in the mass of the vector, 
therefore we keep only the terms that will contribute to the Maxwell vector propagator. 
We have in Einstein frame 
\be
e^{-1}{\cal L}_{Max.} = -\frac{\alpha + 16 \xi Y^2}{4} F^2  - \frac{v^2}{6 \left( Z+ \frac12 \right)^2} \, , 
\ee
after integrating out $H_a$. 
Then after we perform the Weyl rescaling and take the large $Z$ limit (large $\phi$) where quintessence takes place we find 
\be
\alpha + 16 \xi Y^2  \to 4 \xi^{1/3} \e^{\frac{2}{3} \sqrt{\frac23} \phi}  
\ , \quad  
3 \left( Z+ \frac12 \right)^2  \to \frac34 \e^{2 \sqrt{\frac23} \phi} \, . 
\ee
Finally we find for the canonically normalized vector the mass 
\be
m_{v-\text{can.}}^2 \sim \e^{-2 \sqrt{\frac23} \phi / M_P} m_\phi^2 \sim \e^{-2 \sqrt{\frac23} \phi / M_P} f_{SUSY}^2 / M_P^2 \, , 
\ee
up to some numerical coefficients. 
The generic prediction of this setup is therefore a very light massive Maxwell gauge field.

\section*{Acknowledgments} 

This work is supported by the KU Leuven C1 grant ZKD1118C16/16/005.

\end{document}